\documentstyle[12pt]{article}

\font\twlvrm=cmr12

\newcounter{sectionc}\newcounter{subsectionc}%
\newcounter{subsubsectionc}
\renewcommand{\section}[1] {\vspace{0.75cm}%
\addtocounter{sectionc}{1}
\setcounter{subsectionc}{0}%
\setcounter{subsubsectionc}{0}\noindent
			{\large\bf\thesectionc. \hspace{0.1cm}#1}\par}
\renewcommand{\subsection}[1] {\vspace*{0.2cm}%
\addtocounter{subsectionc}{1}
			\setcounter{subsubsectionc}{0}\noindent
			{\normalsize\bf\thesectionc.\thesubsectionc %
			\hspace{0.3cm} #1}\par}
\renewcommand{\subsubsection}[1]
{\vspace*{0.2cm}\addtocounter{subsubsectionc}{1}
			\noindent{\normalsize\sl\thesectionc.%
			\thesubsectionc.\thesubsubsectionc
			\hspace{0.3cm}#1}\par}
\newcommand{\nonumsection}[1] {\vspace{0.6cm}%
\noindent{\normalsize\bf #1}}

\renewenvironment{thebibliography}[1]
			{\begin{list}{\arabic{enumi}.}
			{\usecounter{enumi}\setlength{\parsep}{14pt}
\setlength{\leftmargin 0.25in}{\rightmargin 0pt}
			\setlength{\itemsep}{0pt} \settowidth
 			{\labelwidth}{#1.}\sloppy}}{\end{list}}

\begin{document}
\twlvrm
%
% *********************************************
%      TITLE and AUTHOR
% *********************************************
%
\renewcommand{\thefootnote}{\fnsymbol{footnote}}
\noindent
\begin{center}
{\large \bf Instantons For Black Hole Pair Production}\footnote[1]{To be published
in a festschrift for J. V. Narlikar, Kluwer Academic Publishers, 1999.}

\vspace{0.25in}
{Paul M. Branoff\footnote[2]{email: branoff@glue.umd.edu} and Dieter R. Brill%
\footnote[3]{email:  brill@physics.umd.edu}}\\
\vspace{1ex}
{\it The University of Maryland}\\
{\it Department of Physics}\\
{\it College Park, Maryland 20740}
\end{center}

\vspace{0.15in}
%
% *********************************************
%     ABSTRACT
% *********************************************
%
\begin{abstract}
We address the issue of constructing continuous
instantons representing the pair creation of black holes in a cosmological
context.  The recent attempt at constructing such solutions using
virtual domain walls is reviewed first.  We then explore the existence of
continous instantons in higher curavature gravity theories where
the Lagrangian is polynomial in the Ricci scalar.  Lastly, we study
continous instanton solutions of ordinary gravity coupled to 
the Narlikar $C$-field.  For each theory, we first consider the case of finding
continuous instanton solutions which represent the ``near-annihilation'' of
a de~Sitter universe and its subsequent recreation.  In situations where
these solutions exist, we then ask whether solutions can be found that
represent the ``near-annihilation'' of a de~Sitter spacetime and the subsequent
creation of a pair of Schwarzschild-de~Sitter black holes. 
\end{abstract}

\renewcommand{\thefootnote}{\arabic{footnote}}

\section{Introduction}
Of Jayant Narlikar's many important contributions to astrophysics and 
cosmology, none is more creative and imaginative than the idea, developed 
with Fred Hoyle, that particles may be created as the universe expands.
Stated long before quantum effects of gravity could be treated,
this proposal has new meaning today. Methods are now available to
analyze quantum particle production in dynamic spacetimes, and even
black hole creation can be understood semiclassically as a tunneling process.
The latter process is the main subject of this paper.

Although a complete theory of quantum gravity does not yet exist,
examples of gravitational tunneling have been studied for a number of
years, including such processes as
pair creation of black holes and vacuum decay of
domain walls. In each case the treatment is based on an instanton
(solution of the Euclidean field equations)
that connects the states between which tunneling is taking place. 
However, there are some
nucleation processes of interest where the standard instanton method is
not effective, for example because no
solutions exist to the Euclidean Einstein equations that smoothly connect 
the spacelike sections representing the initial
and final states of the tunneling process. It is therefore an
interesting challenge to adapt the ``bounce" method, most
suitable for vacuum decay calculations, to deal with non-static initial states
and background fields such as a positive cosmological
constant or domain walls typically present when
particle-like states are created.

A positive cosmological constant (and other strong gravitational
sources, such as a positive energy density domain wall) acts to 
increase the separation of timelike geodesics. It is therefore
expected to ``pull particles out of the vacuum" by favoring creation of pairs
over their annihilation. The analogous creation of black hole pairs
in de Sitter space can be treated in WKB approximation by the ``no boundary"
realization of quantum cosmology \cite{Hawking1}. The first 
(and usually only) step in
such a treatment consists of finding a solution of the Euclidean field
equations containing the initial state (pure de Sitter universe) and the
final state (Schwarzschild-de Sitter space) as totally geodesic boundaries.
Such a solution exists only if we accept it in two disconnected pieces.
If the cosmological constant is large enough one then obtains an appreciable
probability of creating in each Hubble volume a pair of black holes 
comparable to the volume's size; if these break up into smaller ones
(see, for example, Gregory and Laflamme \cite{Laflamme}) one has,
within pure gravity, a model of continuous creation not too far removed
in spirit from that of Hoyle and Narlikar.

This model is, however, not fully satisfactory in several respects. For
example, it is not clear how to calculate the ``prefactor" of the exponential
in the transition probability, which would define the dimensionful rate of
the process.  When it can be calculated from 
the fluctuations about the instanton \cite{Cole2}, a ``negative mode" is 
necessary for a  non-vanishing rate.
But this negative mode would have to connect the two parts of the instanton,
and therefore cannot be treated as a small perturbation. A discontinuous
instanton is of course also conceptually unsatisfactory, because the usual
composition rules assume that  histories are continuous.

Each of the two parts of the disconnected instanton has the universe's
volume reaching zero. By forbidding arbitrarily small volumes one can
connect the two parts. The exploration of modifications of Einstein 
gravity in which this is possible is still in its infancy.
For example, Bousso and Chamblin \cite{BC98}
have used virtual domain walls to construct interpolating
instantons.  A similar technique using `pseudomanifolds'
has also been used to construct such solutions \cite{FGG}. 

Modifications of Einstein's theory that have been proposed in other
contexts may also give continuous instantons, if the change from
Einstein's theory becomes important at small volumes. For this reason
it is natural to consider higher curvature gravity theories. 

Another promising modification of Einstein gravity is Narlikar's
$C$-field \cite{Nar}. This field can describe reasonable energetics of particle
production in a context where quantum mechanics plays no essential
role, and it is therefore interesting to explore, as we will below,
whether it can also solve the disconnectedness problem in the
instanton treatment.

So we ask whether these modifications of the Einstein-Hilbert action allow
continuous paths from an initial to final cosmological state
when calculating amplitudes for cosmological black hole production 
in the context of closed universes.  We will outline a modified
version of the calculation of Bousso and Chamblin concerning the use of
virtual domain walls in constructing interpolating instantons.
We next discuss the existence of continuous instantons in
higher curvature gravity theories whose Lagrangians are nonlinear in the
Ricci scalar.  Finally, we consider the case of general relativity
with a cosmological constant and a Narlikar $C$-field.

\section{Gravitational Tunneling}
Processes such as black hole pair creation can be analyzed
semi-classically through the use of instanton methods.
One can think of such a process as a tunneling phenomenon.
The initial state consists of a universe with some
background metric and no black holes, and the final state
consists of a universe with two black holes supplementing the
background metric. Classical dynamics is prevented from connecting
the two states by a generalized potential barrier. The quantum process
can ``penetrate" the barrier with some probability, and the
same barrier makes it improbable for the final state, once 
created, to ``annihilate" back to the initial state. In problems that
can be treated by instantons, the non-classical transition from
initial to final state can be described approximately as  
an excursion in imaginary time.  A solution
that goes from the initial state to the final state
and back again is called a bounce solution; an instanton
is a solution which goes from the initial state to
the final state, i.e., half a bounce. In the WKB interpretation the
excursion into imaginary time simply signifies an exponentially
decreasing wavefunction that is large only near configurations contained
in the instanton. In the sum over histories interpretation the
instanton is a saddle point by means of which the propagator is to be
be evaluated.  

The exponential of the instanton's classical Euclidean action is the 
dominant factor in the transition probability, provided it is normalized
so that the action vanishes when there is no transition. That is, we
are really comparing two instantons, one corresponding to the
background alone in which initial and final states are the same, 
and the instanton of the bounce, in which they are different. If the
initial state is static, it is typically approached asymptotically
by the bounce, and therefore the normalization of the action
can be achieved by a suitable surface term. If the initial state is
only momentarily static, as in the case of the de Sitter universe,
we must find the two instantons explicitly and evaluate their actions.
In the context of the disconnected instanton the background instanton
corresponds to two disconnected halves of a 4-sphere: a de Sitter space
fluctuating into nothing and back again. A first test whether
a modification of Einstein's theory can have connected instantons is
therefore to see whether the background instanton can be connected
(Fig.~1).

The rate of processes
like black hole pair creation is calculated by subtracting from the 
action of the bounce, $I^{\rm bc}$, the action corresponding
to the background state, $I^{\rm bg}$.  The pair creation
rate is then given as 
\begin{equation}
	\Gamma = A\exp{[-(I^{\rm bc} - I^{\rm bg})]} \,,
	\label{eq:Gamma1}
\end{equation}
where $A$ is a prefactor, which is typically neglected in most
calculations because it involves fluctuations about the classical
instantons that are difficult to calculate. Without this dimensionful
prefactor one can find the relative transitions to different final
states, but the actual the number of transitions per spacetime volume 
to a given final state can only be estimated, for example as 
1/(instanton four-volume) for finite volume instantons.

The connected background instanton as described above is closely related 
to a Euclidean wormhole, or birth of a baby universe \cite{Gidd}: if the
two parts are connected across a totally geodesic 3-surface, we can, 
according to the usual rules, join a Lorentzian space-time at that surface,
passing back to real time. An instanton with this surface as the final
state would then describe the fluctuation of a large universe into a
small one, with probability comparable to that of the creation of a 
black hole pair. Thus whatever process provides a connected instanton is
likely to lead not only to the pair creation but also to formation of
a baby universe. (In section 5 we will see how the latter can be avoided)

An instanton calculation has been used by a number of authors to
find the pair creation of black holes on various backgrounds (see,
for example, Garfinkle {\sl et al.} \cite{Garf}). The instantons involved
a continuous interpolation between an initial state without black holes
and a final state with a pair of black holes. By contrast,
in cosmological scenarios
where the universe closes but Lorentzian geodesics diverge, 
as in the presence of a positive cosmological constant or a domain wall,
there are Lorentzian solutions to Einstein's
equations with and without black holes (such as de~Sitter
and Schwarzschild-de~Sitter spacetimes, respectively), but there are no 
Euclidean solutions that connect the spacelike sections of
these geometries \cite{BC98}. (For the related case of baby universe
creation the absence of such solutions is understood, for it is
necessary that the Ricci tensor have at least one
negative eigenvalue \cite{Chee}.)

The No-Boundary Proposal of Hartle and Hawking \cite{Hawking1}
can be modified to provide answers in these cases.
The original proposal was designed to eliminate the initial and final 
singularities of cosmological models by obtaining the universe 
as a sum of regular histories, which may include intervals of imaginary 
time. One can think of the Euclidean sector of the dominant history 
as an instanton that  mediates the creation of a (typically totally 
geodesic)  Lorentzian section from nothing.  By
calculating the action corresponding to these instantons, one
can calculate the wave function for this type of universe, i.e.,
\begin{equation}
	\Psi({\cal G}) = e^{-I_{\rm inst}(\cal G)} 
\end{equation}
where $I_{\rm inst}({\cal G}) = {1\over 2}I^{\rm bc}$ is the action
corresponding to a 
saddlepoint
solution of the Euclidean Einstein equations whose only boundary is the
3-dimensional geometry $\cal G$.
The probability measure associated with this universe
is then given by
\begin{equation}
	P = \Psi^*\Psi = e^{-2I_{\rm inst}} 
\end{equation}
To relate the probability measure to the pair creation
rate of black holes given in Eq.~(\ref{eq:Gamma1}) one writes
\begin{equation}
	\Gamma = \frac{P_{\rm bh}}{P_{\rm bg}} = \exp[-(%
	2I^{\rm bh}_{\rm inst} - 2I^{\rm bg}_{\rm inst})] 
	\label{eq:Gamma}
\end{equation}
so the ratio of the probability of a universe with black holes to the
probability of a background universe without black holes is taken to be
also the rate at which an initial cosmological state can decay
into a final cosmological state, that is, the pair creation rate.
In the latter sense the two disconnected instantons together
describe the tunneling process.

Although this formalism allows one to calculate, in principle,
the rates of nucleation processes, there is no well-justified reason why
Eq.~(\ref{eq:Gamma}) should be identified with this
quantity. The straightforward interpretation of the instanton
concerns the probability for one universe to annihilate to nothing and
for a second universe to be nucleated from nothing. This second
universe can either contain a pair of black holes, or it can be
identical to the initial universe, but it retains no ``memory"
of the initial state. It would clearly be preferable to have
a continuous interpolation between the initial and final states.
(This would allow degrees of freedom that interact only weakly with
the dynamics of gravity to act as a memory that survives the pair
creation.) In the following sections we will consider several ways
in which this continuity of spacetime can be achieved, the first of which
involves matter fields that can form virtual domain walls.

\section{Continuous Instantons via Virtual Domain Walls}
In this section, we will consider the method by which the authors
of \cite{BC98} use virtual domain walls to construct
continuous paths between two otherwise disconnected
instantons.  They illustrated the method for the nucleation
of magnetically charged Reissner-Nordstr\"om black holes 
in the presence of a domain wall. We will confine attention to
nucleation of uncharged black holes in a universe with a cosmological
constant. The initial state is the de~Sitter universe and the final
state is the extremal form of a Schwarzschild-de~Sitter universe
known as the Nariai universe \cite{Nari}, which is dictated by the
requirement that the Euclidean solution be non-singular.
To understand virtual domain walls we will need some elementary properties 
of real domain walls. These have been discussed extensively in 
\cite{BC98, VS, CCG, KE, Vilenkin, IS}.

\subsection{Brief Overview of Domain Walls}
A vacuum domain wall is a $(D-2)$-dimensional topological
defect in a $D$-dimensional spacetime that forms as a
result of a field $\phi$ undergoing the spontaneous breaking
of a discrete symmetry.  If we let ${\cal M}$ denote the
manifold of vacuum expectation values of the field $\phi$,
then a necessary condition for a domain wall to form is 
that the vacuum manifold is not connected ($\pi_0({\cal M}) \ne 0$).  
An example of a potential energy function $U(\phi)$ of the field $\phi$
giving rise to domain walls is the double-well potential.

Throughout this section, we will be dealing with
a domain wall in the ``thin-wall" approximation, which means that  
the thickness of the domain wall is negligible compared to 
its other dimensions, and it is homogeneous and isotropic in
its two spacelike dimensions, so that the spatial section
of the wall can be treated as planar, and the spacetime
geometry as reflection symmetric with respect to the wall.

The action of a real scalar field $\phi$, interacting with gravity,
that may form a domain wall is given by
\begin{equation}
	I_{\rm dw} = \int d^4x \sqrt{-g} \, \left[ %
L_{\rm mat} + \frac{R-2\Lambda}{16\pi} \right] 
\label{eq:dwact}
\end{equation}
with matter Lagrangian
\begin{equation}
L_{\rm mat} = -\frac{1}{2} \, g^{\mu \nu} \, %
\partial_\mu \phi \partial_\nu \phi - U(%
\phi) \,
\label{eq:dwlag}
\end{equation}
and stress-energy tensor
\begin{equation}
	T_{\mu\nu} = \partial_\mu \phi \, \partial_\nu \phi - g_{\mu \nu} %
\left[ \frac{1}{2} \, g^{\alpha \beta} %
\partial_\alpha \phi \, \partial_\beta \phi +%
U(\phi) \right] \,.
\label{eq:TmunuGen}
\end{equation}
Here $U(\phi)$ is a potential function with two degenerate minima
$\phi_-$ and $\phi_+$, at which $U = 0$; $g$ is the determinant of 
the 4-metric $g_{\mu\nu}$; and $R$ is the Ricci scalar.  (We have
neglected boundary terms in the action since
the instantons we will be considering are compact
and have no boundary.)

The trace of the Einstein equations (resulting from the
variation of $I_{\rm dw}$ with respect to $g_{\mu\nu}$)
gives
\begin{equation}
\frac{R - 4\Lambda}{8\pi} = g^{\mu\nu} \, \partial_\mu \phi \,%
\partial_\nu \phi + 4U(\phi) \,,
\end{equation}
which can be used to simplify the action (\ref{eq:dwact}) 
when evaluated on a solution:
\begin{equation}
I_{\rm dw} = \int \left[U(\phi)+{\Lambda\over 8\pi}\right] \sqrt{-g} \, d^4x \,.
\label{eq:Idwact}
\end{equation}

The $\phi$-field is essentially constant away from a domain wall,
with $\phi = \phi_-$ on one side and $\phi = \phi_+$ on the other. 
In Gaussian normal coordinates $(\zeta^i,z)$ with the
wall at $z=0$, $\phi$ depends only on $z$, and the field equation for 
$\phi$ implies that $T_{zz}$ of Eq.~(\ref{eq:TmunuGen}) is negligible. 
The rest of the components of the stress-energy tensor differ from zero
only near the wall, where $\phi$ changes rapidly from $\phi_-$ to
$\phi_+$: 
\begin{equation}
	T^\mu_\nu = \sigma \delta(z) \mbox{diag}(1,1,1,0) 
	\label{eq:Tmunu}
\end{equation}
where $\sigma$ can be related via the $\phi$-field equation to the 
$\phi$-potential alone,
\begin{equation}
	\sigma = \int 2U(\phi(z)) dz  \,.
	\label{eq:usigma}
\end{equation}
Thus $\sigma$ is the surface energy density of the wall. For such surface
distributions the Israel matching condition imply that the intrinsic 
geometry $h_{ij}$ of the domain wall is continuous, and that 
the extrinsic curvature jumps according to 
\begin{equation}
	K_{ij}^+ - K_{ij}^- = 4\pi\sigma h_{ij} \,.
\label{eq:Match}
\end{equation}
Here the normal with respect to which $K_{ij}$ is
defined points from the $+$ side of the surface to the $-$ side.
Outside the wall we have the sourceless Einstein equations.

\subsection{Joining Instantons by Domain Walls}
The jump (\ref{eq:Match}) in extrinsic curvature across a domain wall
can be used to join the two parts of a disconnected instanton (Fig.~1)
by ``surgery": We remove a small 4-ball of radius $\eta$ from each instanton.
Their two 3-surface boundaries have the same intrinsic geometry,
and their extrinsic curvatures are proportional to the surface metric.
They can therefore be joined together in such a way as to
satisfy the Israel matching conditions, Eq.~(\ref{eq:Match}), thereby 
inserting a domain wall.  

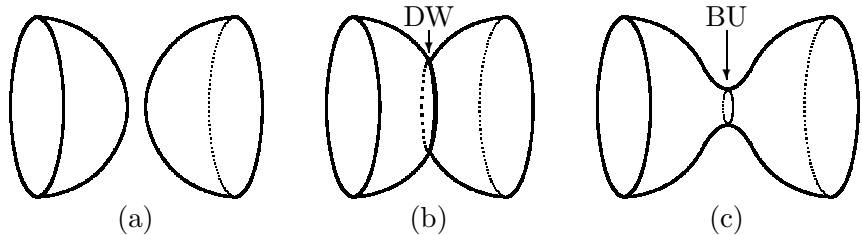
\begin{figure}\small{
\unitlength 0.60mm
\linethickness{0.4pt}
\begin{picture}(202.79,50.00)(-25,0)
\thicklines
\bezier{88}(20.00,50.00)(31.67,49.00)(37.00,40.00)
\bezier{92}(37.00,40.00)(43.00,30.00)(37.00,20.00)
\bezier{88}(37.00,20.00)(31.67,11.00)(20.00,10.00)
\bezier{88}(20.00,50.00)(23.45,49.00)(25.02,40.00)
\bezier{92}(25.02,40.00)(26.79,30.00)(25.02,20.00)
\bezier{88}(25.02,20.00)(23.45,11.00)(20.00,10.00)
\bezier{88}(20.00,50.00)(16.55,49.00)(14.98,40.00)
\bezier{92}(14.98,40.00)(13.21,30.00)(14.98,20.00)
\bezier{88}(14.98,20.00)(16.55,11.00)(20.00,10.00)
\bezier{88}(64.00,50.00)(52.33,49.00)(47.00,40.00)
\bezier{92}(47.00,40.00)(41.00,30.00)(47.00,20.00)
\bezier{88}(47.00,20.00)(52.33,11.00)(64.00,10.00)
\thinlines
\bezier{20}(64.00,50.00)(60.55,49.00)(58.98,40.00)
\bezier{20}(58.98,40.00)(57.21,30.00)(58.98,20.00)
\bezier{20}(58.98,20.00)(60.55,11.00)(64.00,10.00)
\thicklines
\bezier{88}(64.00,50.00)(67.45,49.00)(69.02,40.00)
\bezier{92}(69.02,40.00)(70.79,30.00)(69.02,20.00)
\bezier{88}(69.02,20.00)(67.45,11.00)(64.00,10.00)
\bezier{88}(124.00,50.00)(112.33,49.00)(107.00,40.00)
\bezier{88}(107.00,20.00)(112.33,11.00)(124.00,10.00)
\thinlines
\bezier{20}(124.00,50.00)(120.55,49.00)(118.98,40.00)
\bezier{20}(118.98,40.00)(117.21,30.00)(118.98,20.00)
\bezier{20}(118.98,20.00)(120.55,11.00)(124.00,10.00)
\thicklines
\bezier{88}(124.00,50.00)(127.45,49.00)(129.02,40.00)
\bezier{92}(129.02,40.00)(130.79,30.00)(129.02,20.00)
\bezier{88}(129.02,20.00)(127.45,11.00)(124.00,10.00)
\bezier{88}(90.00,50.00)(101.67,49.00)(107.00,40.00)
\bezier{88}(107.00,20.00)(101.67,11.00)(90.00,10.00)
\bezier{88}(90.00,50.00)(93.45,49.00)(95.02,40.00)
\bezier{92}(95.02,40.00)(96.79,30.00)(95.02,20.00)
\bezier{88}(95.02,20.00)(93.45,11.00)(90.00,10.00)
\bezier{88}(90.00,50.00)(86.55,49.00)(84.98,40.00)
\bezier{92}(84.98,40.00)(83.21,30.00)(84.98,20.00)
\bezier{88}(84.98,20.00)(86.55,11.00)(90.00,10.00)
\bezier{84}(106.90,40.00)(109.26,30.00)(106.90,20.00)
\bezier{84}(107.00,40.00)(109.56,30.00)(107.00,20.00)
\bezier{84}(107.10,40.00)(109.86,30.00)(107.10,20.00)
\put(42.00,5.00){\makebox(0,0)[cc]{(a)}}
\put(107.00,5.00){\makebox(0,0)[cc]{(b)}}
\bezier{10}(107.00,40.00)(105.67,41.33)(105.17,30.00)
\bezier{10}(107.00,20.33)(105.67,19.00)(105.17,30.33)
\bezier{88}(150.00,50.00)(161.67,49.00)(167.00,40.00)
\bezier{88}(167.00,20.00)(161.67,11.00)(150.00,10.00)
\bezier{88}(150.00,50.00)(153.45,49.00)(155.02,40.00)
\bezier{92}(155.02,40.00)(156.79,30.00)(155.02,20.00)
\bezier{88}(155.02,20.00)(153.45,11.00)(150.00,10.00)
\bezier{88}(150.00,50.00)(146.55,49.00)(144.98,40.00)
\bezier{92}(144.98,40.00)(143.21,30.00)(144.98,20.00)
\bezier{88}(144.98,20.00)(146.55,11.00)(150.00,10.00)
\bezier{88}(196.00,50.00)(184.33,49.00)(179.00,40.00)
\bezier{88}(179.00,20.00)(184.33,11.00)(196.00,10.00)
\thinlines
\bezier{20}(196.00,50.00)(192.55,49.00)(190.98,40.00)
\bezier{20}(190.98,40.00)(189.21,30.00)(190.98,20.00)
\bezier{20}(190.98,20.00)(192.55,11.00)(196.00,10.00)
\thicklines
\bezier{88}(196.00,50.00)(199.45,49.00)(201.02,40.00)
\bezier{92}(201.02,40.00)(202.79,30.00)(201.02,20.00)
\bezier{88}(201.02,20.00)(199.45,11.00)(196.00,10.00)
\put(173.00,5.00){\makebox(0,0)[cc]{(c)}}
\bezier{108}(167.00,40.00)(173.00,28.00)(179.00,40.00)
\bezier{108}(167.00,20.00)(173.00,32.00)(179.00,20.00)
\thinlines
\bezier{5}(173.19,33.81)(172.53,33.62)(172.23,31.90)
\bezier{5}(172.23,31.90)(171.90,30.00)(172.23,28.10)
\bezier{5}(172.23,28.10)(172.53,26.38)(173.19,26.19)
\bezier{22}(173.19,33.81)(173.85,33.62)(174.15,31.90)
\bezier{23}(174.15,31.90)(174.48,30.00)(174.15,28.10)
\bezier{22}(174.15,28.10)(173.85,26.38)(173.19,26.19)
\put(173.00,47.00){\vector(0,-1){11.00}}
\put(173.00,48.00){\makebox(0,0)[cb]{BU}}
\put(107.00,47.00){\vector(0,-1){5.00}}
\put(107.00,48.00){\makebox(0,0)[cb]{DW}}
\end{picture}}

\caption{Two-dimensional analog of de Sitter instanton. Imaginary time runs 
horizontally. Because no significant change can be shown in two dimensions, 
this is a ``background" instanton with identical initial and final states.  
(a) The disconnected instanton.  (b) ``Yoyo" instanton connected by domain wall 
(heavy curve labeled DW). (c) Instanton connected by a ``virtual baby 
universe" (BU).}
\end{figure}

However, the surface energy density $\tilde{\sigma}$ of the 
domain wall used to join the instanton must be negative:  
As we approach the domain wall from the initial state, 
heading towards annihilation, successive
3-spheres are shrinking, $K_{ij}^+ < 0$. After we pass through
the domain wall, successive 3-spheres are expanding, $K_{ij}^- > 0$. 
Because of the negative energy density the authors of \cite{BC98} call
this a virtual domain wall, but it is not virtual in the sense that it
corresponds to a Euclidean solution of the equations of section 3.1, for
the $\sigma$ of Eq.~(\ref{eq:usigma}) remains positive when passing to
imaginary time. Within this scheme the only way to achieve a ``yoyo"
instanton as a saddle point of the Euclidean action is to have a scalar 
field with a negative energy also in the real domain, that is, a Lagrangian
with the opposite sign as that of Eq.~(\ref{eq:dwlag}). As we will see in 
section 5, in that case a plain scalar field, without the domain-wall-forming
potential $U(\phi)$, will do as well and is preferable.

By how much does the Euclidean action change when we 
introduce a domain wall whose radius $\eta$ is small compared to
the radius $\sqrt{3/\Lambda}$ of the instanton itself? 
The extrinsic curvature of the connecting 3-sphere is then nearly the same 
as what it would be in flat space,
$K_{ij} = h_{ij}/\eta$, and the jump in curvature is twice that; hence the
size of the domain wall is determined from Eq.~(\ref{eq:Match}),
\begin{equation}
	\eta = -\frac{1}{2\pi \tilde{\sigma}}\,.
\end{equation}
The Euclidean version of Eq.~(\ref{eq:Idwact}) is
\begin{equation}
I_{\rm dw} = - \int \left[U(\phi)+{\Lambda\over 8\pi}\right] \sqrt{g} \, d^4x \,.
\label{eq:Edwact}
\end{equation}
We have taken 
a 4-ball with scalar curvature $R = 4\Lambda$
away from each part of the original instanton, for a total change in
action by $\pi\Lambda \eta^4/16$; this is small compared to that due to the 
added domain wall with action given by Eqs.~(\ref{eq:Edwact}) and 
(\ref{eq:usigma}), 
$I_{\rm dw}=-\pi^2\tilde{\sigma}\eta^3 ={1\over 2}\pi \eta^2 = 
{1\over 64\pi}\times{\eta^2\over 3/\Lambda}\times {96\pi^2\over \Lambda}$, 
which is small compared to the total action $96\pi^2/\Lambda$. Thus the 
Euclidean action increases when we add the domain wall,
and the connected instanton therefore has a relatively smaller
probability measure (although the difference is 
small compared to the total action), and the disconnected instanton
will dominate. If the path integral is extended over continuous histories
only, the domain wall provides the only
saddle point, with action very close to what the discontinuous history
would have given, thus justifying the calculation using the discontinuous
history alone. But in that case a path integral without a domain-wall-forming
scalar field gives a very similar value of the action, as shown in 
\cite{BC98}.

Introducing this scalar field may therefore be considered a high price to
pay for gaining a saddle point, particularly because it entails other,
less desirable processes. For example, the ``center" $z=0$ of the domain wall
is totally geodesic with $\partial\phi/\partial z = 0$, that is, a possible
place to revert from imaginary time back to real time. This corresponds
to the formation of a baby universe of size comparable to $\eta$ and 
smaller Euclidean action than that for the black hole formation.

If a field exists that can form small domain walls, any two instanton
parts can be connected by such surgery across one or several small 3-spheres, 
with a change in action as estimated above for each; the
dominant history will have the fewest connections. 

Finally, recall that the periodicity in imaginary time of each part of the
disconnected instanton is well defined by the requirement that conical 
singularities should be absent from each part. If the parts are connected
where there would otherwise be a conical singularity, one such requirement
is eliminated. Thus there are connected instantons for which the
final state is not Nariai but Schwarzschild-de~Sitter geometry with
black hole and cosmological horizons of unequal size.

\section{Continuous Instantons in Higher Curvature Theories}

Higher curvature theories have a long history and have been proposed 
in several different contexts. For example, they arise naturally
in theories describing 
gravity by an effective action \cite{Myers, Deser}.

In this section we will explore whether higher order theories can
pass the ``first test" of Section 2, namely whether there is a
continuous instanton describing the annihilation and
rebirth of de~Sitter space (generalized to these theories).
Adding higher order terms to the action does not, however, immediately
eliminate disconnected instantons; for example, de~Sitter space (that is,
a spacetime of constant curvature) is a solution of many higher-order
theories. In fact, if the universe without and with black holes can
originate by tunneling from nothing, a disconnected instanton will also
exist. Therefore connected instantons
may again co-exist with the de~Sitter-like, disconnected instantons.

The Euclidean action we will be considering has the form
\begin{equation}
	I = -\frac{1}{16\pi} \int f(R) \sqrt{g} \, d\,^4x
	\label{eq:If(R)}
\end{equation}
where
\begin{equation}
	f(R) = R - 2\Lambda + \alpha R^2 + \gamma R^3 + \cdots \,,
	\label{series}
\end{equation}
$R$ is the Ricci scalar, $\Lambda$ is the cosmological
constant, and $\alpha$, $\gamma$, etc., are coupling
constants whose value we leave unspecified for the moment.
The metric has the Euclidean Robertson-Walker form appropriate
to three-dimensional space slices of constant positive 
curvature:\footnote{The cases
of zero or negative curvature present additional normalization 
problems because the naive Euclidean action would be infinite. 
Therefore we confine attention to the positive curvature case.}
\begin{equation}
	ds^2 = N^2(\tau)d\tau^2 + a^2(\tau)d\Omega^2_3 \,.
	\label{RW}
\end{equation}
Here $\tau$ is imaginary time determined from the analytic continuation
$t \rightarrow i\tau$, $N$ is the lapse function, $a$ is the universe
radius and $d\Omega^2_3$ is the metric on the unit three-sphere. 
Having the metric depend on $N$ and $a$ allows us to obtain all the
independent Einstein equations by varying only these functions in
the action (\ref{eq:If(R)}): variation with respect to $a$ gives us 
the one independent spacelike time development equation, 
and variation with respect to $N$ yields the timelike 
constraint equation, as in ordinary Einstein theory.
A further variation that is easily performed is a {\em conformal} change
of the metric, giving the trace of the field equations, which is not
independent of the other equations but involves only the function $f(R)$:
\begin{eqnarray}
	\frac{\partial}{\partial a} (fNa^3) - \frac{d}%
	{d \tau} \left[ \frac{\partial}{\partial \dot{a}}%
	(fNa^3) \right] + \frac{d^2}{d \tau^2}%
	\left[\frac{\partial}{\partial \ddot{a}}(fNa^3) \right] & = & 0 
	\quad\label{eq:nangang} \\
	a^3f + \frac{\partial f}{\partial N}Na^3 - %
	\frac{d}{d \tau} \left[ \frac{\partial}%
	{\partial \dot{N}} (fNa^3) \right] & = & 0 %
	\label{eq:ntautau} \\
	2Rf^{\prime} + 6\nabla^2f^{\prime} - 4f & = & 0 %
	\label{eq:ntrace}
\end{eqnarray}
where
\begin{equation}
	\nabla^2 = \frac{d^2}{d \tau^2} + %
	\frac{3\dot{a}}{a}\frac{d}{d \tau} \,,
\end{equation}
a dot denotes $d/d\tau$, and a prime denotes $d/dR$.

Equation (\ref{eq:nangang}) is a fourth order ordinary differential
equation, and Eq.~(\ref{eq:ntautau}) is a third order first integral 
of this equation. The trace equation (\ref{eq:ntrace}) shows that we can 
regard $R$ as an independent variable, satisfying a second order equation. 
In this view Eq.~(\ref{eq:ntrace}) replaces Eq.~(\ref{eq:nangang})
(to which it is equivalent), and $a$ also satisfies a second-order
differential equation, namely its definition in terms of $R$,
\begin{equation}
	R = -6\left( \frac{\ddot{a}}{aN^2} - \frac{\dot{a}\dot{N}}%
	{aN^3} + \frac{\dot{a}^2}{a^2N^2} - \frac{1}{a^2} \right) \,.
	\label{defR}
\end{equation}
In addition we still have the constraint,
Eq.~(\ref{eq:ntautau}), a first order relation between $a$ and $R$. 

A general Hamiltonian analysis (c.f. \cite{Buch} and references therein),
not confined to the symmetry of Eq.~(\ref{RW}), bears out the idea that,
as a second-order field theory,
this is Einstein theory coupled to a non-standard scalar field
\cite{Barr}.  For
example, for a Lagrangian quadratic in the Ricci scalar
with no cosmological constant, the relationship between $R$ and the
non-standard scalar field $\phi$ is given by \cite{HawkLut} 
\begin{equation}
	\phi = \sqrt{\frac{3}{4\pi}}\alpha R 
\end{equation}
where the $\phi$-field has the standard stress energy tensor
multiplied by $(1+4(\pi/3)^{1/2}\phi)^{-2}$.

Can this effective scalar field form a domain wall in four dimensions?
If the coefficients up to $\gamma$ in Eq.~(\ref{series}) are non-zero,
then the ``force term" $Rf'-2f$ occurring in the trace equation 
(\ref{eq:ntrace}) vanishes at three equilibrium
states for $R$, where $\nabla^2f'(R)=0$, approximately at $R = 
\pm 1/\sqrt{\gamma}$ and at $R=4\Lambda$, for small $\gamma$. But in
order to have a macroscopic universe on either side of the wall
we need $R=4\Lambda$ on either side, so the usual wall formation where
the scalar field changes from one equilibrium to another is unsuitable
in this case. A solution of the bounce type may appear possible,
since the equilibrium at $R=4\Lambda$ is unstable. At the turning point
the time-dependent $R$ would then have to ``overshoot"
the stable equilibrium near $R=-1/\sqrt{\gamma}$.
A negative $R$ is required there so that the universe radius
can turn around at the same moment. This synchronization, if possible at
all (numerical calculations have failed to reveal it to us; see however
ref.~\cite{Bert}), would require
fine tuning that does not appear natural in this context. 
Furthermore, if we had a bounce for both $a$ and $R$, 
half of it would be an instanton describing the formation of a baby 
universe of size $\sim \gamma^{-1/4}$, which would
then continue to collapse classically, and this process would be
exponentially more probably than the black hole formation. For
these reasons the effective scalar field that derives from higher
curvature Lagrangians of the form (\ref{eq:If(R)}) does not appear
promising for connected instantons.  

We therefore consider solutions to Eqs~(\ref{eq:nangang}) - 
(\ref{eq:ntrace}) when $R$ is constant, $R = R_0$. To allow a
continuous transition to imaginary time at $\tau=0$ we make the
usual ansatz that all odd time
derivatives of $a(\tau)$ vanish at $\tau = 0$. With the choice
$N=1$, the above equations at $\tau = 0$ take the form
\begin{eqnarray}
	R_0f^{\prime} - 2f & = & 0 \\
	a_0f + 6\ddot{a}_0f^{\prime} & = & 0 \\
	a^2_0f - 2f^{\prime}(a_0\ddot{a}_0-2) & = & 0\,.
\end{eqnarray}
Eliminating $f$ from these equations we get the condition
\begin{equation}
	(a^2_0R_0 - 12)f^{\prime} = 0 \,.
\end{equation}
Thus, we have two classes of solutions.  The first class is
described by the condition $R_0 = 12/a_0^2$.  The second
class is described by the condition $f=f^\prime = 0$.

If $R = R_0 = 12/a_0^2$ and $\dot{a}_0 = 0$ then the unique
regular solution of Eq.~(\ref{defR}) is a de~Sitter-like solution, 
$a(\tau) = a_0 \cos (\tau/a_0)$, leading only to the disconnected
instanton.

The second condition indeed allows periodic,
non-collapsing solutions with any amplitude $A$ of the form 
\begin{equation}
	a^2 = {6\over R_0}+ A\cos\sqrt{R_0\over 3}\tau 
	\qquad {\rm where} \qquad f(R_0) = 0 = f'(R_0)\,.
\end{equation}
If we want this $R_0$ to be close to that of the de~Sitter universe, 
$R_0=4\Lambda$, then at least one of the higher-order coefficients 
($\alpha,\;\gamma\;...)$ in $f(R)$ of Eq.~({\ref{series}) has to
be large and rather fine tuned. Furthermore, because the action for all
of these solutions vanishes, we should integrate over all values of $A$,
which includes some disconnected instantons, so this problem is not really 
avoided by these solutions. (They appear pathological also in other ways, for 
example they would allow production of baby universes of any radius.
They would also tend to be unstable in the Lorentzian sector,
although this can be confined to the largest scale by judiciously
choosing $f(R) = R-2\Lambda$ except near $R_0\sim 2\Lambda$.)

\section{Continuous Instantons in $C$-field Theory}

Except for boundary terms, which describe classical matter creation and
which we neglect in the present context, the $C$-field Lagrangian is 
similar to the usual scalar field Lagrangian without self-interaction
(Eq.~(\ref{eq:dwlag}) with $U=0$), but with the important difference
that the coupling constant $-f$ of the $C$-field has the opposite sign from 
the usual one \cite{Nar}. Thus the total action of gravity with cosmological
constant and $C$-field has the form, for Lorentzian geometries
\begin{equation}
	I_{C} = \int d^4x \sqrt{-g} \, \left[\frac{1}{2}f \, g^{\mu \nu} \,
\partial_\mu C \partial_\nu C + \frac{R-2\Lambda}{16\pi} \right]\,. 
\label{CLag}
\end{equation}
(We have not included ordinary matter fields here because we are confining
attention to pair production of black holes as purely geometrical objects.)

\subsection{Sourceless $C$-field in Lorentzian Cosmology}
The field equations that follow from this action by varying $C$ and $g_{\mu\nu}$
are, for the $C$-field:
\begin{equation}
\Box C = C^\mu_{\;;\mu} = 0 \quad ({\rm where}\; C_\mu = C_{,\mu})
\label{Ceq}
\end{equation}
and for the geometry,
\begin{equation}
G_{\mu\nu}+\Lambda g_{\mu\nu} = T^C_{\mu\nu} \quad {\rm where}\quad
T^C_{\mu\nu} = -f\left(C_\mu C_\nu - 
{1\over 2} g_{\mu\nu} C^\alpha C_\alpha\right)\,.
\end{equation}
The stress-energy tensor $T^C_{\mu\nu}$ gives a negative energy density
(for $f > 0$). Narlikar \cite{Nar} has given reasons why 
this violation of the energy condition is not an
objection when the $C$-field is coupled to Einstein gravity of an expanding
universe.

For Lorentzian cosmology we make a Robertson-Walker ansatz analogous to 
(\ref{RW}),
\begin{equation}
	ds^2 = -N^2(t)dt^2 + a^2(t)d\Omega^2_{3} \,.
	\label{LRW}
\end{equation}
In agreement with the homogeneous nature of this geometry we assume that
$C$ is homogeneous in space and hence depends only on $t$. The field 
equations, derived by varying $a$, $N$, and $C$, and then setting $N=1$, are 
\begin{eqnarray}
2{\ddot{a}\over a} + {\dot{a}^2 + 1\over a^2} - \Lambda &=& 4\pi f \dot{C}^2 
\label{varya}\\
\dot{a}^2+1-\frac{\Lambda}{3} a^2 &=& -{4\pi f\over 3}\, \dot{C}^2 a^2 \\
{d(a^3\dot{C})\over a^3\, dt} &=& 0\, .
\end{eqnarray}

The second equation, as usual, is a first integral of the first (time
development) equation, and it implies the latter except for extraneous
solutions $a =$ const. The third equation has the integral
\begin{equation}
	\dot{C} = {K\over a^3}
	\label{dotC}
\end{equation}
where $K$ is a constant. By eliminating $\dot{C}$ we obtain an equation of
the ``conservation of energy" type for $a$:
\begin{equation}
\dot{a}^2+V_{\rm eff} =\dot{a}^2 - \frac{\Lambda}{3} a^2 + {4\pi f K^2\over 
3a^4} = -1 \, . 
\label{Veff}
\end{equation}
This is the usual de~Sitter equation supplemented by a term in $1/a^4$,
which is unimportant at late times when $a$ is large and 
does not change the qualitative Lorentzian time development at any time
(Fig.~2).

\begin{figure}
\unitlength 1mm
\linethickness{0.4pt}
\begin{picture}(61.00,51.00)(-50,5)
\thicklines
\bezier{212}(10.00,30.00)(29.67,30.00)(55.00,10.00)
\bezier{292}(54.00,10.00)(18.89,38.00)(17.00,10.00)
\bezier{64}(55.00,11.00)(47.50,16.50)(42.00,20.00)
\bezier{96}(17.00,49.00)(20.00,35.00)(27.00,30.00)
\bezier{72}(27.00,30.00)(31.50,26.50)(42.00,20.00)
\thinlines
\put(10.00,51.00){\makebox(0,0)[cb]{$V_{\rm eff}$}}
\put(9.00,30.00){\makebox(0,0)[rc]{$0$}}
\put(9.00,20.00){\makebox(0,0)[rc]{$-1$}}
\put(22.00,28.00){\makebox(0,0)[rt]{a}}
\put(15.00,47.00){\makebox(0,0)[lb]{b}}
\put(15.00,14.00){\makebox(0,0)[lt]{c}}
\put(61.00,30.00){\makebox(0,0)[lc]{$a$}}
\put(10.00,30.00){\vector(1,0){50.00}}
\put(10.00,10.00){\vector(0,1){40.00}}
\bezier{100}(10.00,20.00)(35.33,20.00)(60.00,20.00)
\end{picture}

\caption{The effective potential for de~Sitter-like universes: a universe
with only cosmological constant (curve a), one with a real $C$-field
(curve b), and one with a virtual $C$-field (curve c).}
\end{figure}
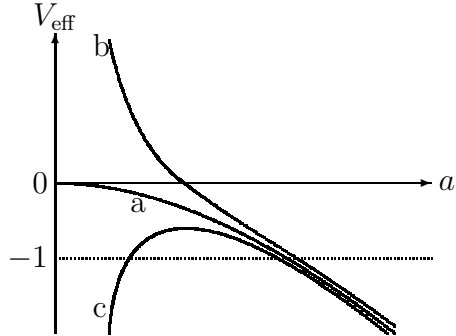

\subsection{Sourceless $C$-field in Euclidean Cosmology}
The effective potential in Eq.~(\ref{Veff}) increases monotonically
as $a$ decreases below the minimum classically allowed value. The
corresponding Euclidean motion in such a potential therefore does not
bounce; instead, $a$ would continue to decrease and reach $a=0$ in a 
finite Euclidean time. This is a geometrical singularity 
if $K\neq 0$ because, for example, it follows from Eqs.~(\ref{varya})
and (\ref{dotC}) that $R = 4\Lambda+8\pi f(K^2/a^6)$.

However, a different potential is obtained if the motion of both $a$
and $C$ is continued to imaginary time,\footnote{We assume that 
this transition is the most probable; this would 
not be so if a transition were possible in the potential of Eq.~(\ref{Veff}).
For example in penetrating radially a spherically symmetric 
potential barrier the most likely transition maintains the real
angular momentum \cite{KLee}.} thereby describing a virtual process
that involves both of these variables, so that we take
into account fluctuations in $C$ as well as in $a$.
Then the $K$ of
Eq.~(\ref{dotC}) becomes imaginary, $K = ik$, and the Euclidean
``conservation of energy" equation becomes
\begin{equation}
-\dot{a}^2 - \frac{\Lambda}{3} a^2 - {4\pi f k^2\over 3a^4} = -1 \, . 
\label{}
\end{equation}
It is easily seen that, for sufficiently small $k$, 
this equation does have bounce solutions, with a turning point 
at $a \sim k^{1/2}f^{1/4}$ (Fig.~2). Thus the $C$-field theory
passes the ``first test": it has a continuous instanton describing a
fluctuation with identical initial and final state. It is reasonable to 
suppose that the theory will also have continuous instantons describing
the creation of a black hole pair, because for small $k$ the
turning point occurs at small $a$, so that two disconnected
instantons can be joined by surgery similar to that of section 3.

It is essential that the fluctuation of the $C$-field be virtual, that is,
that the coupling constant $f$ have the opposite sign from the usual,
positive energy density scalar field. If the $C$-field were real,
time could revert to real values at the minimum radius of the bounce
and continue in a small, Lorentzian universe \cite{Brown1} 
that we have above described as a baby universe.
This transition would be the most probable if allowed. By contrast, 
in the case of the virtual $C$-field this transition is not allowed,
The reason is that at the moment of the bounce, the $C$-field's 
effective potential dominates.
A return to real time ($K$ changing from imaginary to real) would make a large
change in $V_{\rm eff}$ of Eq.~(\ref{Veff}), violating this Lorentzian 
Hamiltonian constraint. A much smaller violation is involved at the
first change to imaginary time, at large $a$. This can occur if the
background is not exactly de~Sitter-like, but contains some gravitational
wave excitation that can supply the necessary small energy difference in the
local region where the black hole will form. Thus the $C$-field makes
a continuous instanton possible, but avoids forming a baby
universe.\footnote{We also note that, as remarked in \cite{Cot}, a real
change in $C$ (if $K$ were real) during the instanton could be interpreted 
as a change in the gravitational constant after the pair creation, which
would be undesirable.} 

\subsection{Black Holes in C-field Cosmology}
As a final step we exhibit as an endstate of the particle creation instanton
an expanding universe in $C$-field theory of spatial topology $S^1\times S^2$.
This describes a universe with an extremal black hole pair in the same sense
that the Nariai solution \cite{Nari, Bousso} describes such a universe
in Einstein's theory. The metric has the homogeneous form
\begin{equation}
	ds^2 = -dt^2 + a^2(t)d\chi^2 + b^2(t)(d\theta^2 + \sin^2\theta d\phi^2)
\end{equation}
where $\chi$ has periodicity appropriate to $S^1$, $\theta$ and $\phi$ are
coordinates on $S^2$, and $a$ and $b$ are functions only of $t$.
The $C$-field likewise is a
function only of $t$ and therefore obeys the conservation law analogous to
(\ref{dotC}),
\begin{equation}
	\dot{C} = {K\over ab^2}\,.
	\label{Cdot}
\end{equation}
The field equations then take the form
\begin{eqnarray}
      G_t^t + \Lambda =  
	-{2\dot a \dot b\over ab}- \frac{\dot{b}^2+1}{b^2} + \Lambda & = &%
		\frac{4\pi f K^2}{a^2 b^4} \\
      G_\chi^\chi + \Lambda =  
	-{2\ddot b\over b}- \frac{\dot{b}^2+1}{b^2} +\Lambda & = & -\frac{4\pi f K^2}{a^2 b^4}\\
	G_\theta^\theta + \Lambda =
	-{\ddot a\over a} - {\dot a\dot b\over ab} - \frac{\ddot{b}}{b} + \Lambda &=&%
		-\frac{4\pi f K^2}{a^2 b^4}\,. 
\end{eqnarray}
If the universe volume expands similar to the Nariai solution, the effects
the $C$-field will become negligible at late times. It is therefore reasonable
to solve the field equations with the condition that the solution
be asymptotic to the Nariai universe, $a(t) =(1/\sqrt{\Lambda})\cosh\sqrt{\Lambda}t$,
$b(t)=1/\sqrt{\Lambda}$. We also require a moment of time-symmetry (to enable the transition from
imaginary time). The solution to first order in $\varepsilon = 4\pi f K^2 \Lambda^{3/2}$ is
\begin{eqnarray}
	a(t) & = &{1\over\sqrt{\Lambda}}\cosh\sqrt{\Lambda}t - \frac{\varepsilon}{3} 
\ln(2\cosh\sqrt{\Lambda}t)- {\varepsilon\over 8}e^{-2\sqrt{\Lambda}t}+\cdots \\ 
	b(t) & = & {1\over\sqrt{\Lambda}} +{\varepsilon\over 6}e^{-2\sqrt{\Lambda}t}+ \cdots\,. 
\end{eqnarray}
These functions do not differ much from those for the Nariai solution for any
time $t$. However, the differences would become large in the continuation to
imaginary time, as the volume decreases. In order to reach a minimum volume
we again need an imaginary $K$ (virtual $C$-field). This minimum volume, like
all $t =$ const. surfaces, has topology $S^1\times S^2$ and would therefore
not fit directly on the minimum-$a$ surface of a de~Sitter-like metric,
Eq.~(\ref{RW}); a solution with
less symmetry in both spaces would be needed to make the match.

\section{Conclusions}
In Einstein's theory of gravity 
with a cosmological constant, typical Euclidean solutions describe
a universe originating from ``nothing," or decaying into nothing, but there
are no equally simple solutions corresponding to quantum processes, 
such as creation of a pair of black holes, which change a universe that 
is already present. According to the simple interpretation of Euclidean
solutions in Einstein's theory, the most probable path to black hole 
creation is discontinuous via nothing as an intermediate state. 
In the present paper we have considered several modifications of Einstein's 
theory that allow continuous histories as saddle points of the Euclidean 
action between two finite universes. Considered as a matter source, these
modifications involve extreme forms of the stress-energy tensor because
the Ricci tensor will typically have at least one negative eigenvalue.
Therefore the formation of baby universes is a possible competing process. 

A matter field that can form sufficiently small domain walls is a universal
connector, replacing the intermediate state of nothing with at least a
small three-sphere. Higher-order Lagrangians in the scalar curvature
have to be fine tuned to allow the desired continuous histories. In many
ways the most successful solution involves a scalar $C$-field of
negative (but small) coupling constant.

%
% ***********************
%     REFERENCES
% ***********************
%
\nonumsection{References}

\end{document}